\documentclass[%
preprint,
 amsmath,amssymb,
 aps,
pre,
]{revtex4-2}

\usepackage{graphicx}
\usepackage{dcolumn}
\usepackage{bm}

\def\C{\mathbb{C}}
\def\R{\mathbb{R}}

\def\Z{\mathbb{Z}}

\def\ro{\mathrm{o}}
\def\rS{\mathrm{S}}
\def\rI{\mathrm{I}}
\def\rR{\mathrm{R}}

\def\bP{\mathbf{P}}

\def\cG{\mathcal{G}}
\def\cN{\mathcal{N}}

\def\bra{\langle}
\def\ket{\rangle}

\begin{document}


\title{Spreading and Suppression of Infection Clusters \\
on the Ginibre Continuum Percolation Clusters}

\author{Machiko Katori}
\email{katori-machiko@g.ecc.u-tokyo.ac.jp}
\affiliation{%
Department of Mathematical Informatics, 
Graduate School of Information Science and Technology, 
The University of Tokyo, 
Hongo, Bunkyo-ku, Tokyo 113-0033, Japan
}%


\author{Makoto Katori}
\email{katori@phys.chuo-u.ac.jp}
\affiliation{
Department of Physics, Faculty of Science and Engineering, 
Chuo University, 
Kasuga, Bunkyo-ku, Tokyo 112-8551, Japan
}%


\date{\today}
             
\begin{abstract}
Off-lattice SIR models are constructed
on continuum percolation clusters
defined on the Ginibre point process (GPP)
and on the Poisson point process (PPP).
The static percolation transitions in 
the continuum percolation models
as well as the infection-spreading transitions
in the SIR models, which are regarded as
the dynamic percolation transitions,
are enhanced in the GPP-based model
compared with the PPP-based model.
This enhancement is caused by
hyperuniformity of the GPP.
On the other hand, 
in the extinction phase of infection
on the phase diagram,  
a wide parameter region is determined
in which formation of infection clusters
is suppressed in the GPP-based model
compared with the PPP-based model. 
We think that the PPP approximates a disordered configuration
of individuals and the GPP does
a configuration of citizens adopting a strategy to keep
social distancing in a city in order to avoid contagion.
The suppression of infection clusters 
observed in the GPP-based model
implies that such a strategy 
is effective when the infectivity is relatively small.
\end{abstract}

\maketitle


\clearpage
\section
{Introduction} \label{sec:introduction}

In the previous paper \cite{KK21}, we 
introduced a new type of 
stochastic epidemic models defined on two-dimensional 
continuum percolation clusters.
For the continuum percolation model
known as the \textit{standard Gilbert disk model} \cite{Gilb61,MR96,BR06}, 
we prepare a random configuration of points on $\R^2$
and put a set of disks with the same radius $r$
centered at the points. 
When the distances of points are less than $2r$, 
the disks centered at these points overlap and
form clumps of disks called \textit{percolation clusters}. 
A statistical ensemble of random point configurations
specified by a probability law is
generally called a \textit{point process} \cite{DVJ03}. 
Here we assume that the point process has translational symmetry 
with a constant finite density $\rho$. 
The product of $\rho$ and the area of a disk $a=\pi r^2$,
\begin{equation}
\kappa := \rho a = \rho \pi r^2, 
\label{eqn:kappa}
\end{equation}
is called the \textit{filling factor}.
A unique \textit{critical value}
$\kappa_{\rm c}$ is defined such that
if $\kappa \leq \kappa_{\rm c}$
any percolation cluster includes only finite number of
disks with probability one, while
if $\kappa > \kappa_{\rm c}$ we have a positive
probability $\Theta(\kappa)>0$ to observe
a percolation cluster involving an infinite
number of disks an (\textit{infinite percolation cluster}). 
We interpret each sample of point process
as a realization of spatial configuration of individuals, 
and if and only if two disks overlap, then 
the individuals living at the centers of these disks
are regarded as neighbors of each other. 
Infection can occur only between neighboring individuals,
and hence our model is for a contagious disease. 
We introduce a parameter $\lambda$ indicating
\textit{infectivity} so that each susceptible (S) individual
will be infected with the rate $\lambda$ multiplied
by the total number of neighboring infected (I) individuals.
And each I-individual becomes recovered (R) with 
a constant rate which is normalized to be unity.
That is, we consider a SIR model with parameter $\lambda$ 
on percolation clusters specified by 
a given point process with filling factor $\kappa$.
When $\kappa > \kappa_{\rm c}$, we can find
an infinite percolation cluster with probability $\Theta(\kappa)>0$
as mentioned above, 
and hence if we consider the SIR model on 
an infinite percolation cluster,
we will see a phase transition at a critical infectivity
$\lambda_{\rm c}(\kappa)$
between an \textit{extinction phase} 
$(\lambda \leq \lambda_{\rm c}(\kappa)$)
and an \textit{infection-spreading phase}
$(\lambda > \lambda_{\rm c}(\kappa)$).
In the former phase any infection process
ceases with a finite duration of time with probability one, 
while in the latter phase processes can continue forever
on an infinite percolation cluster
with a positive probability $\Theta^{\rm SIR}(\kappa, \lambda) >0$.
We notice that such \textit{dynamical phase transitions}
have been extensively studied for 
lattice SIR models 
\cite{Gra83,MDVZ99,dST10,TZ10,SMDHB20,SAAMF20}
(or the lattice SIS models called the \textit{contact processes}
\cite{Lig85,Lig99,MD99,Hin00}).
The phase transitions in the lattice SIR models belong to the \textit{dynamic percolation universality class}
\cite{Gra83,MDVZ99,CG85,Zif21}. 
The present epidemic models are off-lattice generalizations
of the lattice SIR models. 

For an underlying point process of our epidemic model,
we have chosen the \textit{Ginibre point process} (GPP) \cite{Gin65,Shi06,HKPV09}
which has been extensively studied in random matrix theory 
\cite{Meh04,For10,Kat16}. 
We compared infection processes on the GPP-based SIR model
with those on the SIR model defined on the 
\textit{Poisson point process} (PPP) \cite{DVJ03}.
While there is no correlation among points in the PPP,
a repulsive interaction acts between any pair of points
in the GPP and the system becomes 
\textit{hyperuniform} \cite{Tor18,MKS21}.
The hyperuniformity implies \textit{rigidity} of the GPP 
in the sense that occasional appearance of clumping of points
and vacant spaces are suppressed compared with the PPP \cite{GL17}.
As a result, a static percolation transition is
enhanced in the GPP-based continuum percolation model
than in the PPP-based model and hence 
we have an in equality
\begin{equation}
\kappa_{\rm c}^{\rm GPP} < \kappa_{\rm c}^{\rm PPP}.
\label{eqn:ineq_kappa_c}
\end{equation}
Moreover, with a fixed $\kappa_0$ which was chosen to be greater than
$\kappa_{\rm c}^{\rm PPP}$, we observed
$\lambda_{\rm c}^{\rm GPP}(\kappa_0) 
< \lambda_{\rm c}^{\rm PPP}(\kappa_0)$,
which means an enhancement of the dynamic percolation transition to
the infection-spreading phase in the GPP-based SIR model
compared with the PPP-based model \cite{KK21}.

At the same time, we studied infection processes 
with relatively small values of infectivity 
$\lambda < \lambda_{\rm c}^{\rm GPP}(\kappa_0)$ and 
found a parameter regime 
$0 < \lambda < \lambda_*(\kappa_0)$ in which
formation of infection clusters are suppressed in the
GPP-based SIR model compared with the PPP-based model.
We think that the PPP approximates a disordered configuration
of individuals and the GPP does
a configuration of citizens adopting a strategy to keep
social distancing in a city in order to avoid contagion.
The above results seem to prove effectiveness of such a strategy
at least in the time periods between 
epidemic waves of a contagious disease \cite{KK21}.

\begin{figure}[ht]
\includegraphics[width=1\linewidth]{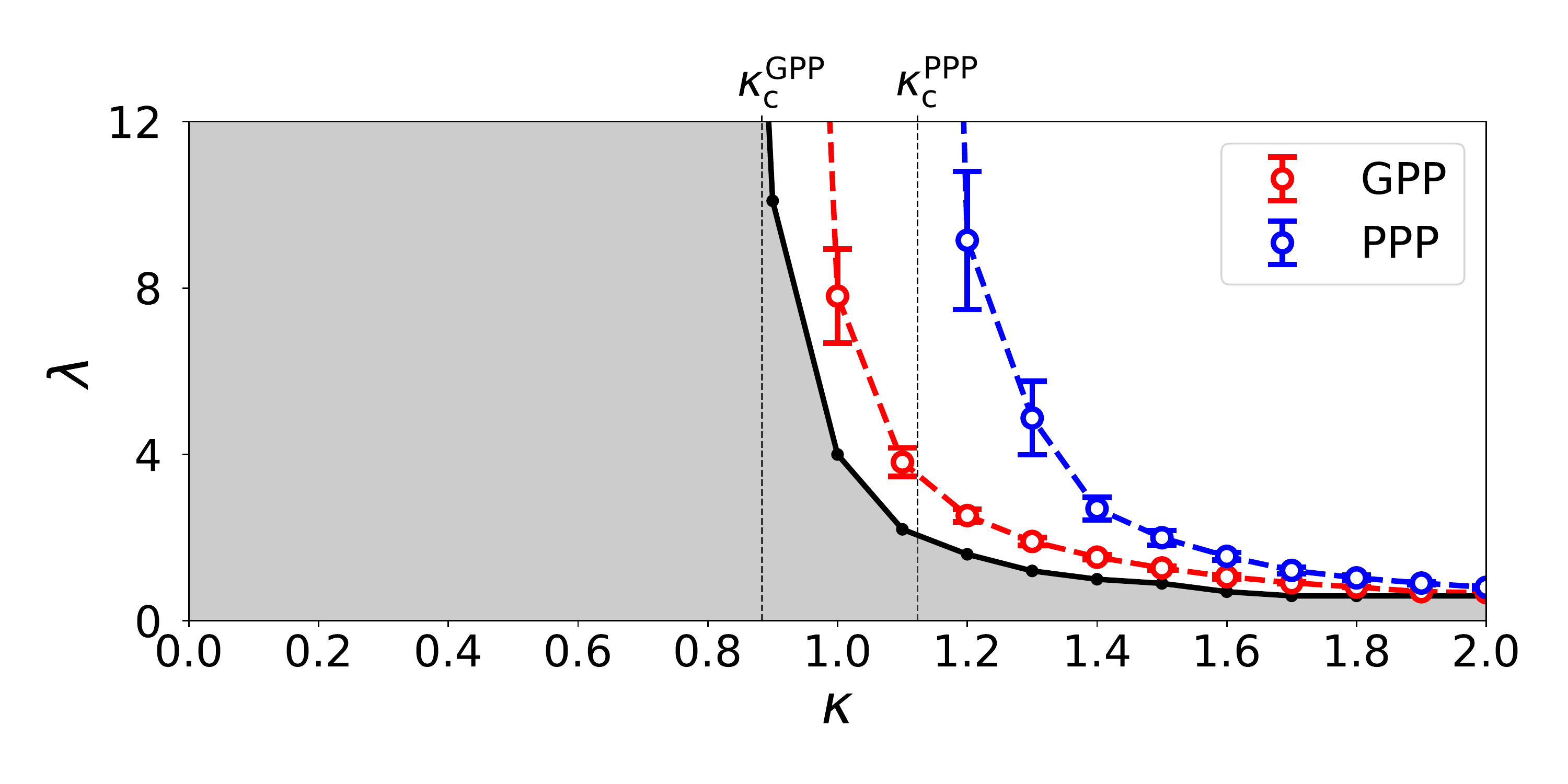}
\caption{
Phase diagram of the present epidemic models 
estimated by numerical simulations.
The phase boundaries between the extinction phases
and the infection-spreading phases are shown by 
a red curve for the GPP-based SIR model
and by a blue curve for the PPP-based model, respectively.
The values of critical filling factors of
continuum percolation transitions 
are indicated by vertical dotted lines, which are
evaluated as $\kappa_{\rm c}^{\rm GPP} \simeq 0.884$
and $\kappa_{\rm c}^{\rm PPP} \simeq 1.12$. 
In the shaded region, infection clusters are suppressed
in the GPP-based SIR model compared with the PPP-based model. 
}
\label{fig:PhaseDiagram}
\end{figure}
\begin{figure}[ht]
\includegraphics[width=1\linewidth]{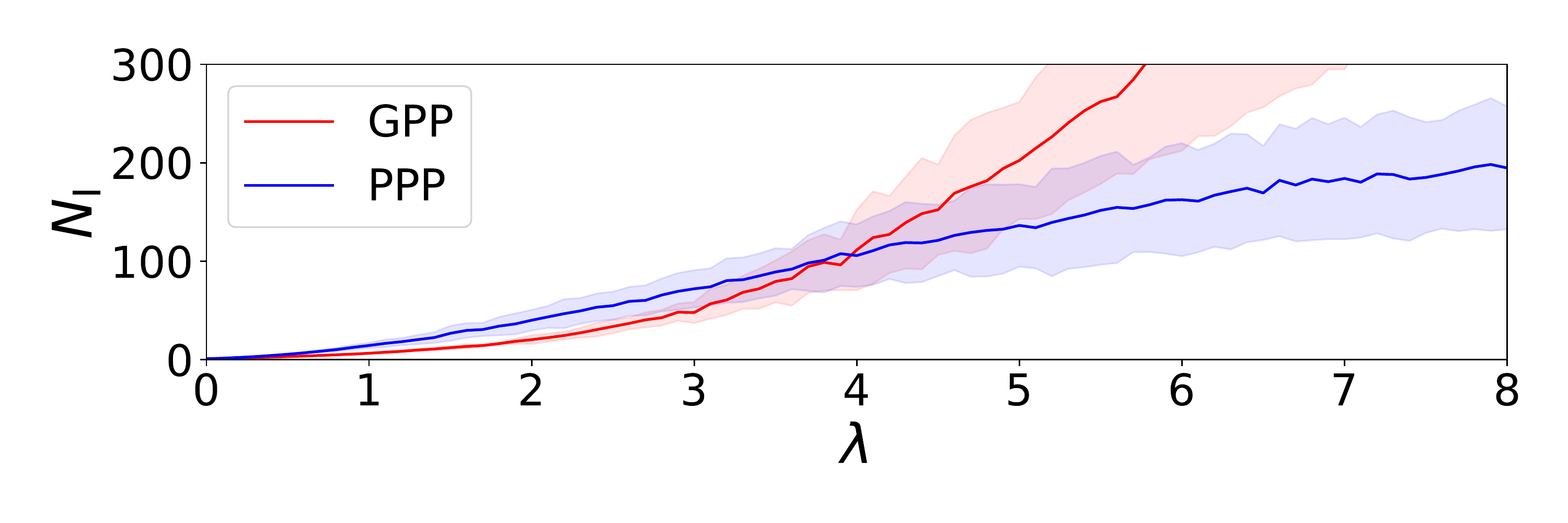}
\caption{
Dependence on $\lambda$ of masses of infection clusters 
in the extinction phase
is shown for the two types of models with $\kappa=1.0$. 
The solid curves show the means
$\bra \cN_{\rm I} \ket$ of 10 samples 
of underlying percolation clusters and
the shaded strips do the standard deviations of them. 
The two curves intersect at $\lambda_* \simeq 4.0$,
which implies that, 
if $\lambda < \lambda_*$
formation of infection clusters is suppressed 
in the GPP-based SIR model compared with
the PPP-based model. 
The curve $\lambda=\lambda_*(\kappa)$,
$\kappa >\kappa_{\rm c}^{\rm GPP}$ 
gives the right boundary of the shaded 
region in Fig.~\ref{fig:PhaseDiagram}. 
}
\label{fig:NI100}
\end{figure}

We have developed our numerical study on these new types of
epidemic models and evaluated the critical values
$\lambda_{\rm c}(\kappa)$ and $\lambda_*(\kappa)$
by systematically changing the filling factor $\kappa$.
In Fig.~\ref{fig:PhaseDiagram}, the estimated phase boundaries between
the extinction phases and the infection-spreading phases,
$\lambda=\lambda_{\rm c}^{\rm GPP}(\kappa)$
and
$\lambda=\lambda_{\rm c}^{\rm PPP}(\kappa)$, 
are plotted.
Both of $\lambda_{\rm c}^{\rm GPP}(\kappa)$ and
$\lambda_{\rm c}^{\rm PPP}(\kappa)$ 
are decreasing in $\kappa$ so that 
for $\sharp =$ GPP and PPP, 
$\lambda_{\rm c}^{\sharp}(\kappa) \to \infty$ as
$\kappa$ is approaching $\kappa_{\rm c}^{\sharp}$ from the above,
$\lambda_{\rm c}^{\sharp}(\kappa) \to 0$
as $\kappa \to \infty$, and the following inequality holds,
\begin{equation}
\lambda_{\rm c}^{\rm GPP}(\kappa) < \lambda_{\rm c}^{\rm PPP}(\kappa),
\quad \kappa > \kappa_{\rm c}^{\rm PPP}.
\label{eqn:ineq_lambda_c}
\end{equation}

In the extinction phase of infection, 
we have focused on the largest, but finite percolation cluster
realized in our simulation and observed 
infection processes on it,
each of which starts from a single infected individual.
We define the mass of infection cluster 
as the total number of
R-individuals in the final configuration of each process.
Since this is equal to the cumulative number of
I-individuals during a process, 
we write it as $\cN_{\rm I}$.
We numerically obtained statistical average
$\bra \cN_I \ket$, which is a function of 
$\kappa$ and $\lambda$.
Figure \ref{fig:NI100} shows the dependence
of $\bra \cN_{\rm I}^{\rm GPP} \ket$ 
and $\bra \cN_{\rm I}^{\rm PPP} \ket$
on $\lambda$ when $\kappa=1.0$. 
We find an intersection 
of the two curves at $\lambda_*(\kappa=1.0) \simeq 4.0$ 
and it implies that, 
if $\lambda < \lambda_*(1.0)$, 
formation of infection clusters is suppressed 
in the GPP-based SIR model compared with
the PPP-based model.
The parameter regime in which 
infection clusters are suppressed 
in the GPP-based SIR model is shown by 
the shaded region in Fig.~\ref{fig:PhaseDiagram}. 

\section
{The Ginibre and the Poisson Continuum Percolation Models} 
\label{sec:percolation}

Let $N \in \{1, 2, \dots\}$. 
We prepared $2N \times 2N$ random matrices
such that the real and the imaginary parts of each entry are
independently and identically distributed
real standard Gaussian random variables.
Then we calculated $2N$ complex eigenvalues and
plot them on the complex plane $\C$
which is identified with the two-dimensional plane $\R^2$. 
We have confirmed that almost all eigenvalues are 
uniformly distributed
in a disk around the origin
with radius $\sqrt{2 N}$;
that is, they follows the \textit{circle law} \cite{Meh04,For10}.
We have used only the point distribution in the
central $\sqrt{\pi N} \times \sqrt{\pi N}$ square
and performed a scale change by factor $1/\sqrt{\pi N}$
to obtain a point process 
on the unit square $[0,1]^2$.
The obtained point process has 
density $\rho^{\rm GPP}_N=N$, 
which well approximates the GPP
and is denoted by $\Xi^{\rm GPP}_N$. 
The PPP is a uniform point process without any correlation
and hence it is easy to generate samples of the PPP 
with $N$ points in $[0,1]^2$.
The obtained point process is denoted by
$\Xi^{\rm PPP}_{N}$ which has
density $\rho^{\rm PPP}_N=N$.
For each sample of point process $\Xi_N$ on $[0, 1]^2$, 
we imposed the periodic boundary condition.

Consider a point process $\Xi$ extended over whole $\R^2$, 
which is a set of an infinite
number of points with a finite constant density $\rho$. 
Then we introduce a real number $r>0$.
We place disks of radius $r$ centered at each point; 
$B_r(x):=\{z \in \R^2 : |z-x|< r\}$,  $x \in \Xi$. 
Two points $x \in \Xi$ and $y \in \Xi$ are neighbors, 
if and only if $B_r(x) \cap B_r(y) \not= \emptyset$.
If there is a finite sequence of points 
$x_0, x_1, \dots, x_n \in \Xi$ such that 
$x_0=x, x_n = y$ and
$x_{j+1}$ is a neighbor of $x_j$, $j=0,1,\dots, n-1$, 
then the two points $x$ and $y$ are \textit{connected}.
The maximal connected components are called
\textit{percolation clusters}
whose sizes are given by 
the numbers of points of $\Xi$ included in the clusters.
The above defines the
\textit{standard Gilbert disk model} with radius $r$,
which is one of 
the \textit{Boolean percolation models}, on the
point process $\Xi$
\cite{Gilb61,MR96,BR06,BY13,BY14,GKP16}. 
It is said that the system percolates
if there is at least one 
\textit{infinite cluster}, 
whose size is infinity \cite{SA92, Ess71}. 
It is proved that an infinite cluster exists uniquely 
with probability one for the Boolean percolation models
on the PPP and the GPP \cite{MR96,BR06,GKP16}. 
The probability that the system percolates is called
the \textit{percolation probability}. 
In the present Boolean percolation model on $\R^2$,
it is a function of the
filling factor $\kappa$ defined by (\ref{eqn:kappa}) \cite{MM12}.
We write the percolation probability as $\Theta(\kappa)$. 
There is a unique \textit{critical value of
filling factor} $\kappa_{\rm c}$ such that
$\Theta(\kappa)=0$ 
if $\kappa \leq \kappa_{\rm c}$,
and $\Theta(\kappa)>0$ if $\kappa > \kappa_{\rm c}$. 

In the standard Gilbert disk model defined
on a finite point process $\Xi_N$ in $[0, 1]^2$, 
$\Theta$ is approximated by the ratio 
$\Theta_N$ of the number of points
included in the largest cluster 
to the total number of points $N$ \cite{SA92}.
We have generated 10 samples of
finite approximations
$\Xi^{\rm PPP}_N$ and $\Xi^{\rm GPP}_N$
for seven different $N$, 
$N=1000, 2000, \cdots, 7000$.
In \cite{KK21}, we performed extrapolation 
of the numerical results to $N \to \infty$
and obtained the following evaluations,
\begin{align}
\kappa_{\rm c}^{\rm GPP} &=0.884 \pm 0.016,
\nonumber\\
\kappa_{\rm c}^{\rm PPP} &=1.12 \pm 0.05.
\label{eqn:kc}
\end{align}
The above estimation
of $\kappa_{\rm c}^{\rm PPP}$
is consistent with the value 
reported in \cite{MM12}
evaluated using efficient algorithms
\cite{QT99,QTZ00,NZ01,QZ07},
$\kappa_{\rm c}^{\rm PPP}=1.12808737(6)$. 

\begin{figure}[ht]
\includegraphics[width=1\linewidth]{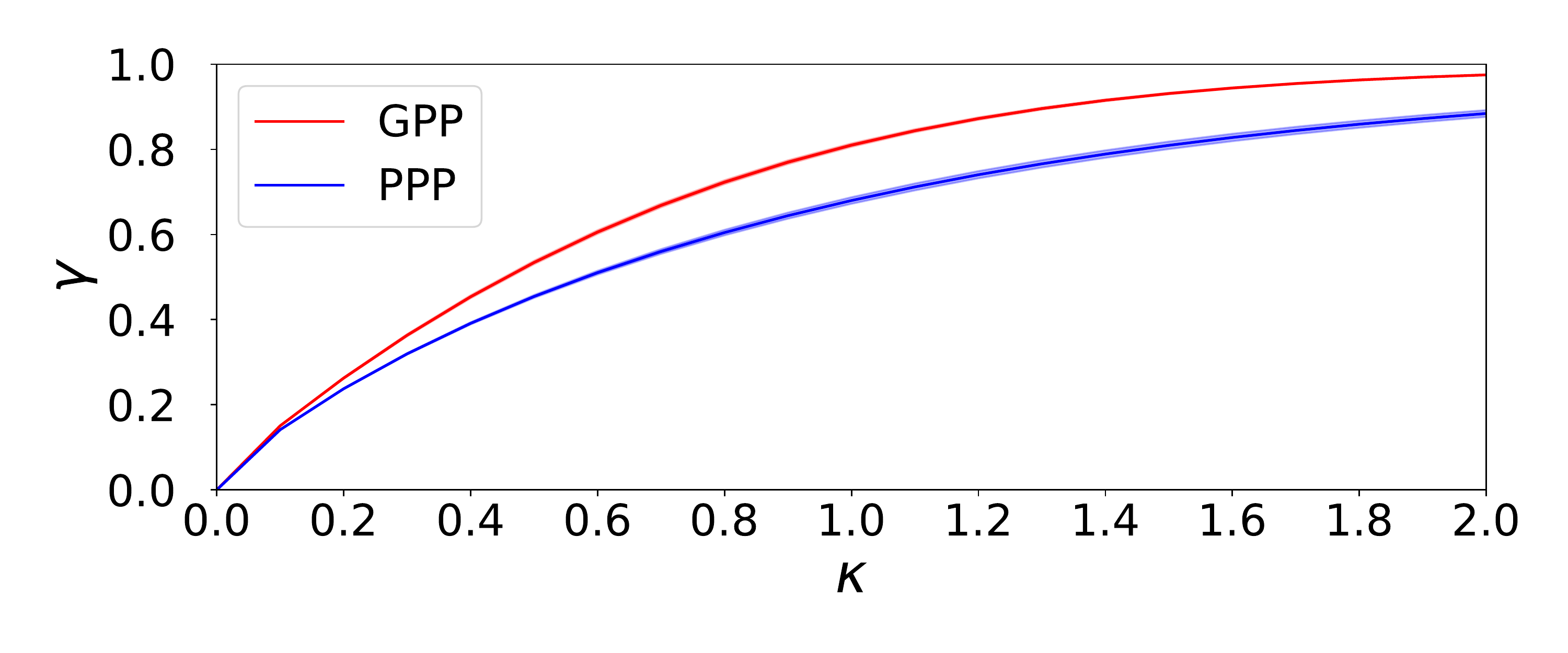}
\caption{
Dependence on $\kappa$ of 
the occupation ratios $\gamma$ of the Ginibre disks
and of the Poisson disks. 
Solid curves show the means of 10 samples
and shaded strips do the standard deviations of them.
}
\label{fig:kappa_gamma}
\end{figure}

By the definition (\ref{eqn:kappa}), $\kappa$ denotes
the total area of disks with radius $r$ 
which are penetrable to each other.
One of the origin of the difference of
$\kappa_{\rm c}^{\rm GPP}$ and $\kappa_{\rm c}^{\rm PPP}$
shown above is the difference of frequencies of
overlapping of disks under the same values of
$\rho$ and $r$ in these two distinct point processes.
Since the GPP is hyperuniform \cite{Tor18,MKS21},
while the PPP is not, the overlapping of disks
occurs less frequently in the GPP than in the PPP
and hence the \textit{net area} of disks will be
larger in the GPP than in the PPP under the same value 
of $\kappa$. 
Therefore, the percolation transition occurs
on the GPP clusters at
$\kappa=\kappa_{\rm c}^{\rm GPP}$ which is
smaller than $\kappa_{\rm c}^{\rm PPP}$.
Here we use the 10 samples with 1000 points
on $[0, 1]^2$ 
for $\Xi_{1000}^{\rm GPP}$ and $\Xi_{1000}^{\rm PPP}$,
in which $\rho_N=N=1000$. 
For each sample of random points on $[0, 1]^2$,
we impose the periodic boundary condition
and increase the radius $r$ of disks centered at
the points. At each value of 
$\kappa=\rho \pi r^2$, we measured 
the area $a_{\rm empty}$ of the subdomain in $[0, 1]^2$
which is \textit{not} covered by any disk at all.
We define the variable $\gamma:=1-a_{\rm empty} \in [0, 1]$,
which indicates the net area of disks
in the unit square,
that is, 
$\gamma=|\bigcup_{x \in \Xi} B_r(x)|$. 
Figure \ref{fig:kappa_gamma} shows 
the occupation ratios of disks
$\gamma=\gamma^{\sharp}(\kappa)$
for $\sharp=$ GPP and PPP.
As expected 
$\gamma^{\rm GPP}(\kappa) > \gamma^{\rm PPP}(\kappa)$
for all $\kappa > 0$. 
By Fig.~\ref{fig:kappa_gamma}, the critical values of
filling factor (\ref{eqn:kc}) are mapped to the following,
\begin{align}
\gamma_{\rm c}^{\rm GPP}=0.763 \pm 0.005,
\nonumber\\
\gamma_{\rm c}^{\rm PPP}=0.719 \pm 0.008. 
\label{eqn:gamma_c}
\end{align}
If we interpret the site percolation model
on the square lattice $\Z^2$ as the Bernoulli distribution
of unit disks on sites with probability $p$,
the percolation threshold 
$p_{\rm c} \simeq 0.5927$ \cite{SA92} will
give the critical occupation ratio
$\gamma_{\rm c}^{\Z^2} 
=(\pi/4) \times p_{\rm c} \simeq 0.4655$. 
The critical values (\ref{eqn:gamma_c})
for the random point processes
are greater than $\gamma_{\rm c}^{\Z^2}$,
which is due to the off-lattice effect.
We see that the relative difference
of $\kappa_{\rm c}$'s,
$|\kappa_{\rm c}^{\rm GPP}-\kappa_{\rm c}^{\rm PPP}|/
(\kappa_{\rm c}^{\rm GPP}+\kappa_{\rm c}^{\rm PPP})
\simeq 0.12$ is reduced to
$|\gamma_{\rm c}^{\rm GPP}-\gamma_{\rm c}^{\rm PPP}|/
(\gamma_{\rm c}^{\rm GPP}+\gamma_{\rm c}^{\rm PPP})
\simeq 0.030$ for $\gamma_{\rm c}$'s. 
It should be noticed that $\gamma_{\rm c}^{\rm GPP}$
is larger than $\gamma_{\rm c}^{\rm PPP}$.
It suggests that even at the percolation threshold,
overlapping of disks occurs less frequently in the 
GPP-based SIR model than in the PPP-based model.
If we interpret centers of disks as 
locations of individuals, the above implies
that the area per one individual in the 
GPP-based SIR model is larger than that
in the PPP-based model.
Hence if the infectivity $\lambda$ is small enough,
infection could be more suppressed in the GPP-based SIR
model compared with in the PPP-based model.

\section
{Spreading of Infection on Continuum Percolation Clusters} 
\label{sec:SIR}

For each finite approximation of point process
$\Xi_N$ on $[0, 1]^2$ with the periodic boundary condition, 
we have defined the continuum percolation
model with the filling factor $\kappa$ as explained above.
Here we consider the supercritical case $\kappa > \kappa_{\rm c}$. 
We pick up the largest percolation cluster 
and write it as $\cG_N$.
The size of $\cG_N$ is defined as the total 
number of disks included in it and denoted by
$|\cG_N|$. 
We use $\cG_N$ as an underlying graph of our contagious
epidemic model of the SIR type as follows.

At each point $x \in \cG_N$, we put a random
variable $\eta(x) \in \{\rS, \rI, \rR\}$.
We consider a continuous-time Markov process,
$(\eta_t)_{t \geq 0}$, where
$\eta_t:=\{\eta_t(x) : x \in \cG_N \}$. 
Let $\bP^{\eta}$ denote
the probability in this Markov process
starting from the configuration 
$\eta :=\{\eta(x): x \in \cG_N \}$. 
The transition mechanism is given by
\begin{align}
\bP^{\eta}(\eta_t(x)=\rI, \eta(x)=\rS)
&=\lambda c(x, \eta) t + \ro(t),
\nonumber\\
\bP^{\eta}(\eta_t(x)=\rR, \eta(x)=\rI)
&=t + \ro(t)
\quad \mbox{as $t \to 0$},
\label{eqn:transition1}
\end{align}
where 
$\lambda$ is a positive parameter called 
the infectivity and 
$c(x, \eta)$ is a positive function 
depending on a point $x$ and a configuration $\eta$.
That is, if an individual at point $x$ is in the
susceptible (S) state, it is infected (I)
at rate $\lambda c(x, \eta)$, 
while if it is infected (I), it becomes recovered (R)
at rate 1. 
Once $\eta(x)=\rR$, the state at $x$ does not change
forever.
We require that only one-variable-change
happens in each transition; that is,
$\bP^{\eta}(\eta_t(x) \not= \eta(x), \eta_t(y) \not= \eta(y))
=\ro(t)$ as $t \to 0$
for each $x, y \in \cG_N$ with
$x \not=y$ given a configuration $\eta$ 
on $\cG_N$ \cite{Lig85,Lig99}. 
Here we set the function $c(x, \eta)$ as 
\begin{equation}
c_{\rm linear}(x, \eta)
=\sum_{y: |y-x| < 2r} 1_{(\eta(y)=\rI)},
\label{eqn:c1}
\end{equation}
where $1_{(\omega)}$ is the indicator function of
an event $\omega$;
$1_{(\omega)}=1$ if $\omega$ is satisfied and
$1_{(\omega)}=0$ otherwise.
In other words, the infection rate is
given by the total number of 
infected neighbors multiplied by 
the infectivity $\lambda$.
In the previous paper \cite{KK21}, 
we called the SIR model with (\ref{eqn:c1}) the \textit{linear model}.
We have simulated this continuum Markov process
on $\cG_N$ by the Gillespie algorithm \cite{Gil76,Gil77,EC20}.
This method is also called the {\it n}-fold way algorithm for spin systems \cite{BKL75}.

If we consider the above SIR model
on an \textit{infinite} percolation cluster $\cG$ 
which is found in the supercritical phase 
$\kappa > \kappa_{\rm c}$ 
formed on an infinite point process,
we can discuss the percolation problem
for an infection cluster consisting of
$\rI$-individuals and $\rR$-individuals
\cite{Gra83,MDVZ99,dST10,TZ10,
SMDHB20,SAAMF20,Lig85,Lig99,MD99,Hin00,CG85,Zif21}.
The percolation probability
of infection cluster denoted by $\Theta^{\rm SIR}$
is increasing in $\lambda$ for each $\kappa>0$ and 
a unique critical value $\lambda_{\rm c}(\kappa)$,
$\kappa > \kappa_{\rm c}$, 
is defined so that
\begin{align*}
&\Theta^{\rm SIR}(\kappa, \lambda)=0, 
\quad \mbox{if $\lambda \leq \lambda_{\rm c}(\kappa)$},
\nonumber\\
&\Theta^{\rm SIR}(\kappa, \lambda) >0, 
\quad \mbox{if $\lambda > \lambda_{\rm c}(\kappa)$},
\quad \mbox{for $\kappa > \kappa_{\rm c}$}.
\end{align*}

In the present numerical simulation, we consider 
the off-lattice SIR model on a finite percolation
cluster $\cG_N$ with $|\cG_N| \leq N$.
Therefore, any infection process on $\cG_N$
ceases sooner or later and in
a final configuration 
we have a percolation cluster consisting of
$\rR$-individuals embedded in a field of
$\rS$-individuals. 
For each $(\kappa, \lambda)$, we simulated the SIR model
from a single infected individual 
100 times and evaluated the mean ratio
of total number of $\rR$-individuals to
$|\cG_N|$ in a final configuration.
We regard this as an approximation
of infection probability $\Theta^{\rm SIR}_N(\kappa, \lambda)$
with size $N$ \cite{Gra83,CG85,dST10,TZ10}.
Increasing the size $N$ of point process $\Xi_N$
systematically, we prepare
a series of approximations
$\{\Theta^{\rm SIR}_N(\kappa, \lambda)\}_{N}$
with $N=1000, 2000, 3000, 4000, 5000, 6000$,
and 7000. 

\begin{figure}[ht]
\includegraphics[width=1\linewidth]{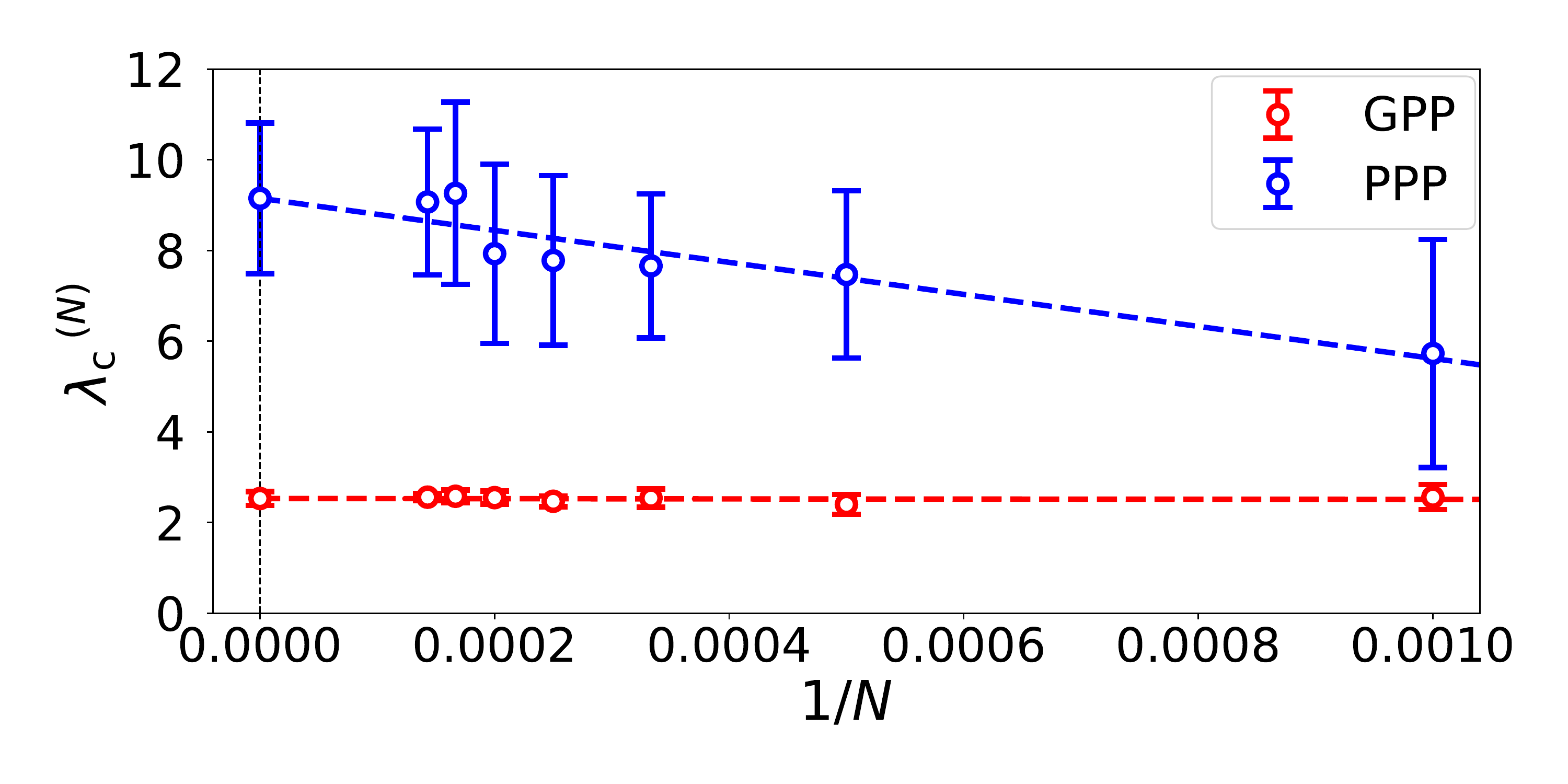}
\caption{
Critical infectivities $\lambda_{\rm c}^{\sharp}(\kappa)$
are evaluated for $\sharp =$ GPP and PPP 
by linear extrapolations of finite-size approximations
$\lambda_{\rm c}^{\sharp (N)}(\kappa)$ 
in the $1/N$-plots. The case with $\kappa=1.2$ is shown.
}
\label{fig:lambda_c}
\end{figure}
We have evaluated $\lambda_{\rm c}(\kappa)$
for $\kappa > \kappa_{\rm c}$ by the following
procedure.
Fix $\kappa > \kappa_{\rm c}$. 
For each $N=1000, \dots, 7000$,
we numerically generated 10 samples of
$\Xi_N$ and then $\cG_N$, 
and have drawn curves $\Theta^{\rm SIR}_N$ versus $\lambda$.
For each curve an approximate value of critical infectivity 
denoted by $\lambda_{\rm c}^{(N)}(\kappa)$ is
defined by the value of $\lambda$ at which
the numerical value of 
$\partial \Theta^{\rm SIR}_N(\kappa, \lambda)/\partial \lambda$ 
attains a maximum.
We plot $\lambda_{\rm c}^{{\rm GPP}(N)}(\kappa)$ and 
$\lambda_{\rm c}^{{\rm PPP}(N)}(\kappa)$
versus $1/N$. 
Figure \ref{fig:lambda_c} shows the plots
for $\kappa=1.2$, where the error bars are estimated
by the 10 samples for each $N$.
The linear extrapolation 
with respect to $1/N$ gives
$\lambda_{\rm c}^{\rm GPP}(1.2)=
\lim_{N \to \infty} \lambda_{\rm c}^{{\rm GPP}(N)}= 2.53 \pm 0.15$,
and
$\lambda_{\rm c}^{\rm PPP}(1.2)=
\lim_{N \to \infty} \lambda_{\rm c}^{{\rm PPP}(N)}=9.15 \pm 1.66$.
By this procedure the phase boundaries between
the extinction phases and the infection-spreading phases
of infection,
$\lambda=\lambda_{\rm c}^{\rm GPP}(\kappa)$,
$\kappa > \kappa_{\rm c}^{\rm GPP}$, 
and
$\lambda=\lambda_{\rm c}^{\rm PPP}(\kappa)$, 
$\kappa > \kappa_{\rm c}^{\rm PPP}$, 
are plotted in Fig.~\ref{fig:PhaseDiagram}. 

\section
{Suppression of Infection Clusters in the GPP-based 
Epidemic Model} 
\label{sec:suppression}

With $N=1000$ we have prepared 10 samples
of $\Xi_N$ and then generate 10 finite percolation clusters
$\cG_N$ with a filling factor $\kappa > 0$.
Here we consider the extinction phase of infection, 
$\lambda < \lambda_{\rm c}(\kappa)$.
On each $\cG_N$, we simulated the present off-lattice SIR model
starting from a single infected individual 100 times.  
For each process of the SIR model
with parameter $\lambda$,
we counted the total number of
R-individuals in the final configuration,
which is equal to the cumulative number of
I-individuals during the process and 
represents the mass of infection cluster.
Then $\cN_{\rm I}$ is defined as its
mean value over the 100 runs of simulations.
Moreover, we have averaged $\cN_{\rm I}$ over 10 
samples of $\cG_N$ to obtain
$\bra \cN_{\rm I} \ket$.
Comparison of 
$\cN_{\rm I}^{\rm GPP}$
and $\cN_{\rm I}^{\rm PPP}$
was already demonstrated by Fig.~\ref{fig:NI100}
for $\kappa=1.0$. 

When $\kappa > \kappa_{\rm c}^{\rm GPP}$, 
we have found an intersection 
of the two curves 
$\bra \cN_{\rm I}^{\rm GPP} \ket=
\bra \cN_{\rm I}^{\rm GPP} \ket(\lambda)$
and
$\bra \cN_{\rm I}^{\rm PPP} \ket=
\bra \cN_{\rm I}^{\rm PPP} \ket(\lambda)$
at $\lambda_*(\kappa)$.
The value of $\lambda_*(\kappa)$ is
decreasing in $\kappa$ and behaves as
$\lambda_*(\kappa) \to \infty$ as
$\kappa$ approaches $\kappa_{\rm c}^{\rm GPP}$ from the above,
and $\lambda_*(\kappa) \to 0$ as
$\kappa \to \infty$.
When $\kappa \leq \kappa_{\rm c}^{\rm GPP}$, 
$\bra \cN_{\rm I}^{\rm GPP} \ket(\lambda)
 < \bra \cN_{\rm I}^{\rm PPP} \ket (\lambda)$
for any $\lambda >0$ and there is no intersecting point.
As shown by the shaded region in the
$\kappa$-$\lambda$ phase diagram 
of the epidemic models (Fig.~\ref{fig:PhaseDiagram}),
formation of infection clusters is suppressed 
in the GPP-based SIR model compared with the PPP-based model
in the parameter regions
(i) $\kappa \leq \kappa_{\rm c}^{\rm GPP}$ and
(ii) $\lambda < \lambda_*(\kappa)$,
$\kappa > \kappa_{\rm c}^{\rm GPP}$.

\section
{Concluding Remarks} 
\label{sec:remarks}

We put three remarks on the present study.

\begin{figure}[ht]
\includegraphics[width=1\linewidth]{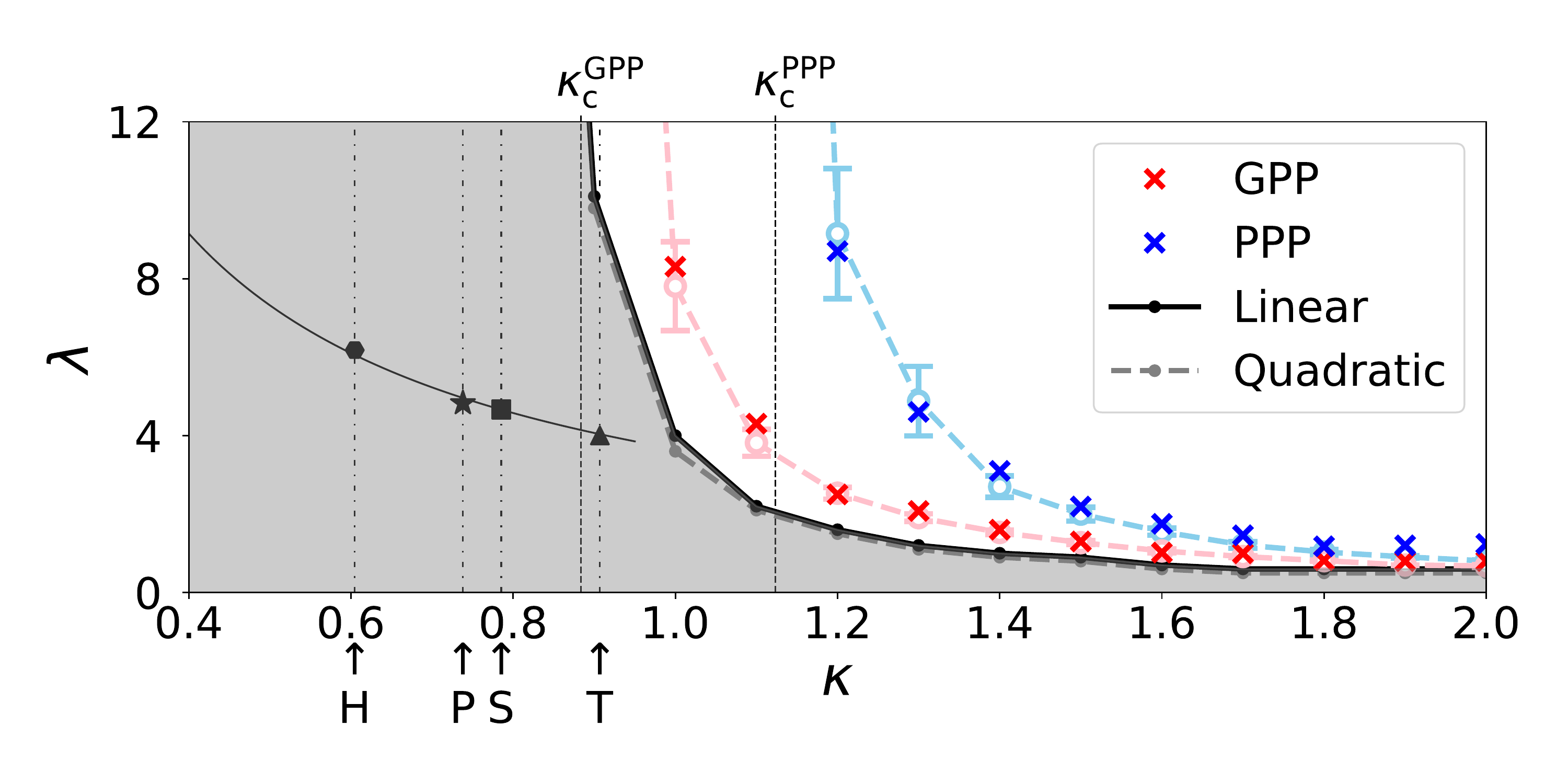}
\caption{
(a) The values of $\lambda_*^{\rm quad}(\kappa)$, 
$\kappa > \kappa_{\rm c}^{\rm GPP}$ 
of the quadratic model are superposed 
on the results of the linear model.
The numerical difference of them is very small.
(b) The critical infectivities $\lambda_{\rm c}(\kappa)$
for $\kappa > \kappa_{\rm c}$ evaluated using
the power law (\ref{eqn:power_law}) are
dotted by red crosses (resp.\ blue crosses) 
for the GPP-based (resp.\ PPP-based) SIR models.
The results are consistent with the phase boundaries
determined by the $1/N$-extrapolation method
of finite-size approximations shown by light red and light blue.
(c) The dependence of $\lambda_{\rm c}$ on $\kappa$ is
shown for the lattice SIR models on the
triangular (T), the square (S), the Penrose (P), 
and the honeycomb (H) lattices,
which is well approximated by
$\lambda_{\rm c} \simeq 3.66/\kappa$.
}
\label{fig:PhaseDiagram2}
\end{figure}

\vspace{3mm}
\noindent(a) \
In the previous paper \cite{KK21}, we studied
variations of the present off-lattice SIR model
by modifying the function
$c(x, \eta)$ specifying the infection transition (\ref{eqn:transition1}).
In particular, we reported that
the suppression of infection clusters
in the GPP-based SIR model is enhanced 
in the \textit{quadratic mode},
in which $c(x, \eta)$ is given by
\begin{equation}
c_{\rm quad}(x, \eta)
= \left( \sum_{y: |y-x| < 2r} 1_{(\eta(y)=\rI)} \right)^2, 
\label{eqn:c2}
\end{equation}
instead of (\ref{eqn:c1}). 
As reported in \cite{KK21}, 
the suppression of mass of infection clusters
$\cN_{\rm I}$ in the GPP-based SIR model
is indeed enhanced in the quadratic model,
but the value of $\lambda_*^{\rm quad}(\kappa)$,
$\kappa > \kappa_{\rm c}^{\rm GPP}$ does not change so much
from $\lambda_*^{\rm linear}(\kappa)$.
See Fig.~\ref{fig:PhaseDiagram2}. 

\begin{figure}[t]
\includegraphics[width=1\linewidth]{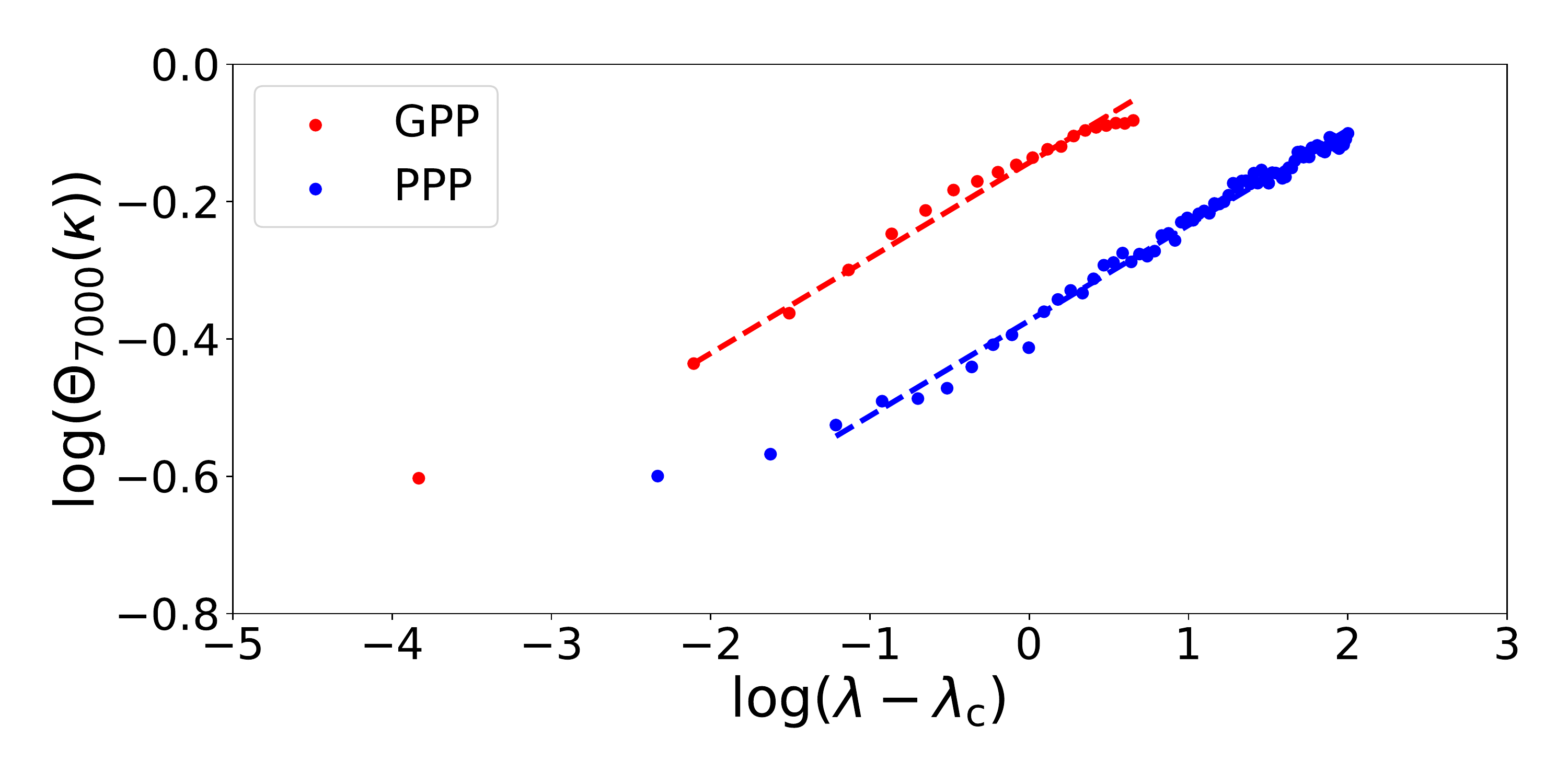}
\caption{
The least-square linear regression was applied to 
the log-log plots of the data
$\Theta^{\rm SIR}_{7000}(\kappa=1.3, \lambda)$ versus
$\lambda-\lambda_{\rm c}(1.3)$, 
where $\lambda_{\rm c}(1.3)$ is a fitting parameter
and the slope of fitting line 
is fixed to be $5/36=0.1388 \cdots$.
The evaluated values are
$\lambda_{\rm c}^{\rm GPP}(1.3) \simeq 2.08$
and $\lambda_{\rm c}^{\rm PPP}(1.3) \simeq 4.60$,
respectively.
}
\label{fig:LogLog}
\end{figure}

\vspace{3mm}
\noindent(b) \
For each $\kappa > \kappa_{\rm c}$, 
we have a critical infectivity $\lambda_{\rm c}(\kappa)$
for the present off-lattice SIR model. 
In the infection-spreading phase
$\lambda > \lambda_{\rm c}(\kappa)$,
$\Theta^{\rm SIR}(\kappa, \lambda)$ is positive
and regarded as the \textit{order parameter}
of dynamic percolation transition.
In the vicinity of the phase boundary, 
it is expected that
$\Theta^{\rm SIR}(\kappa, \lambda)$ 
behaves as
\begin{equation}
\Theta^{\rm SIR}(\kappa, \lambda)
\simeq \mbox{const.}
 (\lambda-\lambda_{\rm c}(\kappa))^{\beta_{\rm SIR}},
\quad \lambda \gtrsim \lambda_{\rm c}(\kappa),
\quad \kappa > \kappa_{\rm c},
\label{eqn:power_law}
\end{equation}
with a critical exponent $\beta_{\rm SIR}$. 
Such a power law indicates one of the critical phenomena.
The present off-lattice SIR model
will belong to the same universality class
with the lattice SIR models,
which is called the \textit{dynamic percolation universality class}, 
and $\beta_{\rm SIR}=5/36$
\cite{Gra83,CG85,MDVZ99,dST10}.
Using the simulation results on the largest systems 
with $N=7000$, 
we applied the least-square linear regression to
log-log plots of the data
$\Theta^{\rm SIR}_{7000}(\kappa, \lambda)$ versus
$\lambda-\lambda_{\rm c}(\kappa)$
for $\kappa > \kappa_{\rm c}$,
where $\lambda_{\rm c}(\kappa)$ is regarded as a fitting parameter,
while the slope of fitting line 
is fixed to be $5/36=0.1388 \cdots$.
Figure \ref{fig:LogLog} shows the best fitting result
when $\kappa=1.3$, which gives 
$\lambda_{\rm c}^{\rm GPP}(1.3)\simeq 2.08$
and $\lambda_{\rm c}^{\rm PPP}(1.3) \simeq 4.60$.
In Fig.~\ref{fig:PhaseDiagram2} we plotted the 
values of $\lambda_{\rm c}^{\sharp}(\kappa)$,
$\sharp=$ GPP and PPP by this method.
The results are consistent with those
obtained by the linear extrapolation method 
$\lambda_{\rm c}^{\sharp}(\kappa)=
\lim_{N \to \infty} \lambda_{\rm c}^{\sharp (N)}(\kappa)$
in the $1/N$-plots explained in 
Section \ref{sec:SIR}.
It means the validity of the assumption (\ref{eqn:power_law}) with $\beta_{\rm SIR}=5/36$.

\vspace{3mm}
\noindent(c) \
While the original SIR model was given by a system of deterministic
differential equations \cite{KM27}, the lattice SIR models
include randomness and are described by stochastic processes \cite{Gra83,MDVZ99,dST10,TZ10,SMDHB20,SAAMF20}.
In the present off-lattice versions,
we have introduced another randomness to generate spatial configurations of individuals. 
It is an interesting problem to compare the lattice SIR models and the present off-lattice SIR models based on random point processes.
As mentioned in the item (b), the additional randomness of underlying graphs seems to be irrelevant for the critical phenomena associated with the infection transitions at $\lambda_{\rm c}$. 
The values of critical infectivity $\lambda_{\rm c}$ are, however, sensitive to the point configurations of individuals. 
For the triangular (T), the square (S), and the honeycomb (H) lattices in two dimensions, the filling factors (1) are calculated as $\kappa^{\rm T}=\sqrt{3} \pi/6=0.906 \cdots$, $\kappa^{\rm S}=\pi/4=0.785 \cdots$, and $\kappa^{\rm H}=\sqrt{3} \pi/9=0.604 \cdots$, respectively.
The critical infectivities were reported as
$\lambda_{\rm c}^{\rm T}=4.0068(2)$ \cite{TSZ13},
$\lambda_{\rm c}^{\rm S}=4.66571(3)$ \cite{TZ10}, 
and $\lambda_{\rm c}^{\rm H}=6.179(5)$ \cite{TSZ13}, respectively.
Recently Santos et al.\ \cite{SAAMF20} simulated the SIR model on the
Penrose tiling and evaluated as
$\lambda_{\rm c}^{\rm P}=(1-0.1713(2))/0.1713(2) = 4.838(7)$.
We have the value
$\kappa^{\rm P}=\sqrt{2} (5-\sqrt{5})^{1/2} \pi/10=0.738 \cdots$ \cite{Hen86}.
As shown by Fig.~\ref{fig:PhaseDiagram2},
we see the tendency such that $\lambda_{\rm c}^\flat$ decreases as
$\kappa^\flat$ increases for $\flat=$ H, P, S, T, which is in common
with the critical lines $\lambda=\lambda_{\rm c}^{\sharp}(\kappa)$,
$\kappa < \kappa_{\rm c}^{\sharp}$ for
$\sharp$ = GPP and PPP. 
Moreover, we notice an approximate reciprocal law 
$\lambda_{\rm c}^\flat \simeq 3.66/\kappa^\flat$,
$\flat =$ H, P, S, T.
The spatial configurations of individuals are deterministic
in the lattice SIR models, random but having hyperuniformity
in the GPP \cite{Tor18, MKS21}, and uniformly random-distributed 
in PPP, respectively.
Hence we say that, for each infectivity $\lambda >0$, the spread of infection can be observed only in higher values of filling factor $\kappa$ of individuals, as the randomness of spatial configuration of individuals is increased.
Further systematic studies of the dependence of $\lambda_{\rm c}$ of the SIR models on the underlying graphs are required, in which deterministic but heterogeneous lattices \cite{SAAMF20}, random point processes \cite{KK21}, as well as a variety of randomly perturbed lattices \cite{GMS20} shall be considered.


\clearpage

\section*{Acknowledgements}
The present authors would like to thank Robert M. Ziff 
for allowing them to use the results reported in their preprint.
They also thank Tomonari Dotera for useful discussion on 
quasicrystal lattices. 
Machiko K. thanks to T.J.Kobayashi for encouraging the present work.
Makoto K. expresses his gratitude to John W. Essam
for a stimulating communication which motivated 
this study.
Machiko K. was supported by the ANRI Fellowship and International Graduate Program of Innovation for Intelligent World (IIW) of The University of Tokyo.
Makoto K. was supported by
the Grant-in-Aid for Scientific Research (C) (No.19K03674),
(B) (No.18H01124), 
(S) (No.16H06338), 
and
(A) (No.21H04432) 
of Japan Society for the Promotion of Science.

\vspace{5mm}
Machiko K. and Makoto K. designed and performed the research and the paper was written cooperatively.
Machiko K. performed computer simulations and data analysis.



\bibliographystyle{bst/apsrev4-2}
\bibliography{MK2-second-reference-re.bib}

\clearpage

\end{document}